\documentclass[11pts, twocolumn]{LML}

\newcommand{\tdm}{\tau_{\rm dm}}
\newcommand{\fdm}{f_{\rm dm}}
\newcommand{\mdm}{m_{\rm dm}}
\begin{document}
\title{ Constraints on ALPs and excited dark matter from the EDGES 21\,cm absorption signal}

\author[a]{Andi Hektor
\thanks{\href{mailto:andi.hektor@cern.ch}{andi.hektor@cern.ch}}
}

\author[a,b]{Gert H\"utsi
\thanks{\href{mailto:gert.hutsi@to.ee}{gert.hutsi@to.ee}}
}

\author[a]{Luca Marzola
\thanks{\href{mailto:luca.marzola@cern.ch}{luca.marzola@cern.ch}}
}

\author[a]{Ville Vaskonen
\thanks{\href{mailto:ville.vaskonen@kbfi.ee}{ville.vaskonen@kbfi.ee}}
}

\affil[a]{\KBFI}
\affil[b]{Tartu Observatory, University of Tartu, Observatooriumi 1, 61602 T\~oravere, Estonia.}
\date{\today}

\twocolumn[

\maketitle

\begin{onecolabstract}
The recent observation of the 21\,cm absorption signal by the EDGES experiment provides a new observational window into the dynamics of the young Universe. Based on this result, we constrain the properties of ALPs and excited dark matter via the energy injections into the gaseous medium that these models produce. In particular, we derive bounds that outperform the present constraints for energy injections in the 10.2~eV to 50~keV range by analysing the intensity of Lyman-$\alpha$ photons and the electron emissivity produced by ionising radiation. Our results also show that once the standard temperature of the soft radiation background is assumed, the explanations of the observed 3.5~keV photon signal within the considered dark matter model are excluded at 95\% confidence level.
\end{onecolabstract}
]
\saythanks

\section*{Introduction} 
\label{sec:Introduction}

The recent detection of the 21\,cm spectrum by the EDGES experiment~\cite{Bowman:2018yin} presents a first insight into the dynamics of the Dark Ages of the Universe, close to the onset of star formation. Within $\Lambda$CDM, the observed absorption feature is explained by the Wouthuysen-Field effect~\cite{1952AJ.....57R..31W, 1958PIRE...46..240F}, after the first stars produced a significant population of Lyman-$\alpha$ photons at $z\sim 20$. The effect of this energetic radiation is then to decouple the hydrogen spin temperature from the soft radiation temperature that characterises the redshifted CMB background, forcing it to converge to the lower kinetic temperature of the gas. 

Whereas the redshift of the observed signal agrees with the theoretical predictions, the measured magnitude of the absorption feature greatly exceeds the $\Lambda$CDM expectations, consequently raising the problem of its origin. Explanations of the EDGES result proposed in literature invoke either new mechanisms for cooling the gaseous medium~\cite{Barkana:2018lgd, Munoz:2018pzp, Berlin:2018sjs, Barkana:2018qrx, Costa:2018aoy, Hill:2018lfx, Slatyer:2018aqg, Falkowski:2018qdj, Li:2018kzs, Lambiase:2018lhs, Houston:2018vrf, Sikivie:2018tml} or the presence of an additional soft photon component~\cite{Feng:2018rje, Ewall-Wice:2018bzf, Fraser:2018acy, Pospelov:2018kdh, Lawson:2018qkc, Moroi:2018vci}. However, the mechanism beyond the anomaly remains to-date unclear. Furthermore, the significance of the measurement and the treatment of the foregrounds is currently under debate~\cite{2018ApJ...858L..10D, Hills:2018vyr}.

In spite of its possible origin, a confirmation of the EDGES measurement would provide a new observable capable of probing the dynamics of the young Universe. As an example, the measured absorption profile strongly constrains the presence of additional mechanisms resulting in Lyman-$\alpha$ radiation at $z\gtrsim 20$, which would otherwise induce earlier absorption signals. At higher energies, the 21\,cm spectrum constrains instead possible energy injections that would lead to ionizations and consequent heating of the gaseous medium. These observations have been used in literature to bound energy injections due to new physics processes~\cite{DAmico:2018sxd, Clark:2018ghm, Cheung:2018vww, Hektor:2018qqw, Liu:2018uzy, Mitridate:2018iag}, the properties of the dark sector~\cite{Safarzadeh:2018hhg, Schneider:2018xba, Lidz:2018fqo} and to investigate the astrophysical consequences implied by the signal~\cite{Mirocha:2018cih, Hirano:2018alc, Witte:2018itc, Sharma:2018agu}.

In this Letter we adopt the same attitude and make use of the EDGES results to investigate the properties of light dark matter (DM) candidates that could result in energy injections in the gaseous medium over the 10.2~eV to 50~keV range, a mass range not considered by the above-cited papers. As concrete examples we consider the case of axion-like particles (ALPs)~\cite{Masso:1995tw} and of excited DM~\cite{Finkbeiner:2007kk}, deriving the phenomenological consequences of the EDGES measurement and comparing the resulting constraints to the ones previously proposed in the literature~\cite{Cadamuro:2011fd, Arias:2012az}. Particular attention is paid to the issue of the 3.5~keV line detected in galaxy clusters~\cite{Bulbul:2014sua, Boyarsky:2014jta}, previously motivated both in excited DM \cite{Finkbeiner:2014sja} and ALPs \cite{Jaeckel:2014qea} models.

Our analyses show that once a temperature of the soft radiation background in agreement with $\Lambda$CDM is assumed, the constraints resulting from the EDGES measurement outperform the ones previously considered in the literature on the considered energy range. Explanations of the observed 3.5~keV photon signal are also excluded at 95\% confidence level for both the examined DM models.

\section*{Energy injection and propagation} 
\label{sec:Formalism}

In the following we provide the basic formalism for describing the impact of decaying or de-exciting DM on the cosmic medium around redshifts $z\sim 20$. By comparing the resulting Lyman-$\alpha$ intensity with the critical intensity initiating the onset the 21\,cm absorption we bound the energy input around 10\,eV. For photons with higher energies we derive the corresponding bound by constraining the heating of the gaseous medium.

\subsection*{Energy input}
The spectral emissivity for a line of energy $E_\star$ can be expressed as follows~\footnote{We denote with a tilde superscript quantities that have dimension of a number density rather than energy density.}

\begin{equation}\label{eq1}
\tilde{\jmath}_E(E,z) = \frac{N}{\tau_{\rm dm}} \frac{f_{\rm dm} \Omega_{\rm dm} \rho_c}{m_{\rm dm}}\, \delta(E-E_*)\, (1+z)^3\,,
\end{equation}

where $N$ is the number of photons with energy $E_*$ emitted per process and $\tdm$ and $\mdm$ are respectively the lifetime and mass of a DM component that constitutes a fraction $\fdm$ of the observed abundance and sources the photon injection.

\subsection*{Lyman-$\alpha$ constraint}

With the above expression for the photon spectral emissivity, we compute the intensity of Ly-$\alpha$ photons as~\cite{Barkana:2004vb}

\begin{equation}\label{eq5}
\tilde{J}_\alpha(z) \hspace{-1mm} = \hspace{-1mm} \sum_{n=2}^\infty \int\limits_z^{z_{\rm max}(n)} \hspace{-2mm}{\rm d}z' \frac{\tilde{\jmath}_E(E'_n,z')}{4\pi H(z') (1+z')} \left(\frac{1+z}{1+z'}\right)^2 \,,
\end{equation}

where

\begin{equation}
	\begin{aligned}
		E_n &= (1-n^{-2})\,{\rm Ry}\,,\\
		E'_n &= \frac{1+z'}{1+z}E_n\,,\\
		1+z_{\rm max}(n) &= (1+z)\frac{E_{n+1}}{E_n}\,,
	\end{aligned}
\end{equation}

with Ry being the Rydberg constant and $n$ the principal quantum number. Notice that due to the monochromatic input spectrum, for a given $z$ only one term contributes to the sum. 

Because for high redshifts $H(z)\simeq H_0\sqrt{\Omega_m}(1+z)^{3/2}$, by using Eq.~\eqref{eq1} Eq.~\eqref{eq5} becomes 

\begin{equation}
\tilde{J}_\alpha(z)= \frac{N}{\tau_{\rm dm}} \frac{f_{\rm dm} \Omega_{\rm dm}\rho_c}{4\pi H_0\sqrt{\Omega_m}}\,\frac{E_n^{1/2}}{m_{\rm dm}E_*^{3/2}}\,(1+z)^{3/2}\,,
\end{equation}

for $E_n<E_*<E_{n+1}$. We can now compare the above quantity to the critical Ly-$\alpha$ intensity necessary to couple the hydrogen spin temperature to the gas kinetic temperature. Because the gas temperature is lower than the radiation one for the redshifts of interest, exceeding this limit would induce a detectable early absorption signal in the 21\,cm spectrum~\cite{Furlanetto:2006jb}. Recent observation by the EDGES experiment~\cite{Bowman:2018yin} require that such signal is absent for $z\gtrsim 20$, hence 

\begin{equation} \label{lacj1}
\tilde{J}_\alpha(z) < 2.3\times 10^{-21}\,(1+z)\,{\rm eV}^2{\rm sr}^{-1}\,.
\end{equation}

A more stringent limit could be derived by considering the opposite end of detected the absorption feature, although this would require to disentangle the contribution of the first stars to the Ly-$\alpha$ intensity.

\subsection*{Gas heating}

Photons with energies larger than the hydrogen ionization energy liberate electrons from the gaseous medium, which in turn heat up the gas via scattering. The electron emissivity is related to that of the ionising radiation via
  
\begin{equation}
\tilde{\jmath}_E^e = \tilde{\jmath}_E \left(1-\frac{\rm Ry}{E_*}\right) \theta(E_*-{\rm Ry})\,,
\end{equation}

where $\theta$ is the Heaviside step function.

\begin{figure}
\begin{center}
\includegraphics[width=0.48\textwidth]{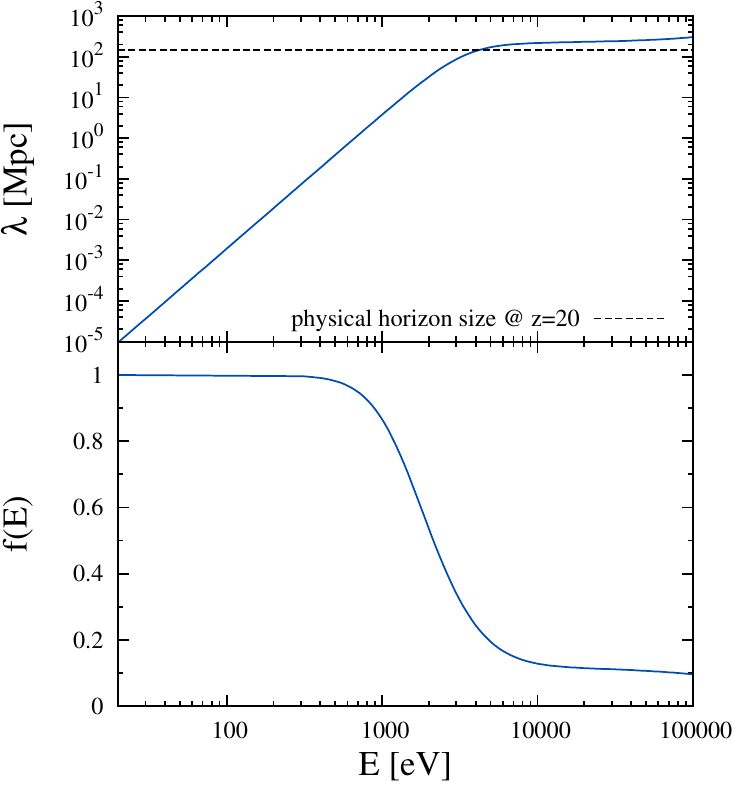}
\end{center}
\caption{The mean free path $\lambda$ and the correction coefficient $f$, which corrects the results of the `on-spot' approximation, as a function of the energy $E$ for $z=20$.}
\label{fig:MFP_f_param}
\end{figure}

To calculate the thermal history of the gas due modified by the above energy input, we adapt the cosmological recombination code RECFAST~\cite{Seager:1999bc} following Ref.~\cite{Padmanabhan:2005es}, neglecting a factor of $(1+z)^3$ in the given definition of ${\cal F}(z)$\footnote{Ref.~\cite{Padmanabhan:2005es} investigated the DM annihilation signal, which scales as $\propto (1+z)^6$.}. The rate of energy injection per hydrogen atom is here given as

\begin{equation}
\epsilon_{\rm dm} = f\,\frac{j^e}{n_{H,0}}\,,
\end{equation}

where the coefficient $f(E)$, shown in figure~\ref{fig:MFP_f_param}, improves on the `on-spot' approximation used in Ref.~\cite{Padmanabhan:2005es} and 

\begin{equation}\label{eq13}
\begin{aligned}
&j^e = \int \tilde{\jmath}_E^eE\,{\rm d}E \\&\hspace{-1mm}= \frac{N}{\tau_{\rm dm}}\frac{f_{\rm dm} \Omega_{\rm dm}\rho_c}{m_{\rm dm}}(E_*-{\rm Ry})\theta(E_*-{\rm Ry})(1+z)^3\,.
\end{aligned}
\end{equation}

As in Ref.~\cite{Padmanabhan:2005es} we assume that a fraction $(1-x_e)/3$ of the energy input results in ionization processes, whereas a share as large as $(1+2x_e)/3$ produces heating of the medium ($x_e$ here is the ionization fraction). The above approximation was suggested in Ref.~\cite{Chen:2003gz}, which itself is based on the earlier work of Ref.~\cite{1985ApJ...298..268S}.

By rewriting Eq.~\eqref{eq13} as $j^e =j^e_0(1+z)^3$ and using this form as an input for thermal history evolution, we obtain the following expression for  the gas kinetic temperature at a redshift $z=17$
 
\begin{equation}
T_k(z=17)\,[{\rm K}]\simeq T_k^{\rm \Lambda CDM} + \kappa j^e_0\,[{\rm eV}^5]\,,
\end{equation}

where $T_k^{\rm \Lambda CDM}\simeq 6.8\,{\rm K}$ is the baseline gas temperature in the $\Lambda$CDM model at $z=17$ and $\kappa \simeq 7.3\times 10^{54}$. Requiring then that $T_k$ does not exceed a limit $T_k^{\rm max}$ leads to the following upper bound

\begin{equation} \label{ghcj1}
j^e_0 \lesssim \frac{T_k^{\rm max} - T_k^{\rm \Lambda CDM}}{\kappa}\,,
\end{equation}

which we adopt to limit the possible electron injection and, therefore, the emissivity of ionizing radiation.

\begin{figure*}[h!]
\begin{center}
\includegraphics[height=0.45\textwidth]{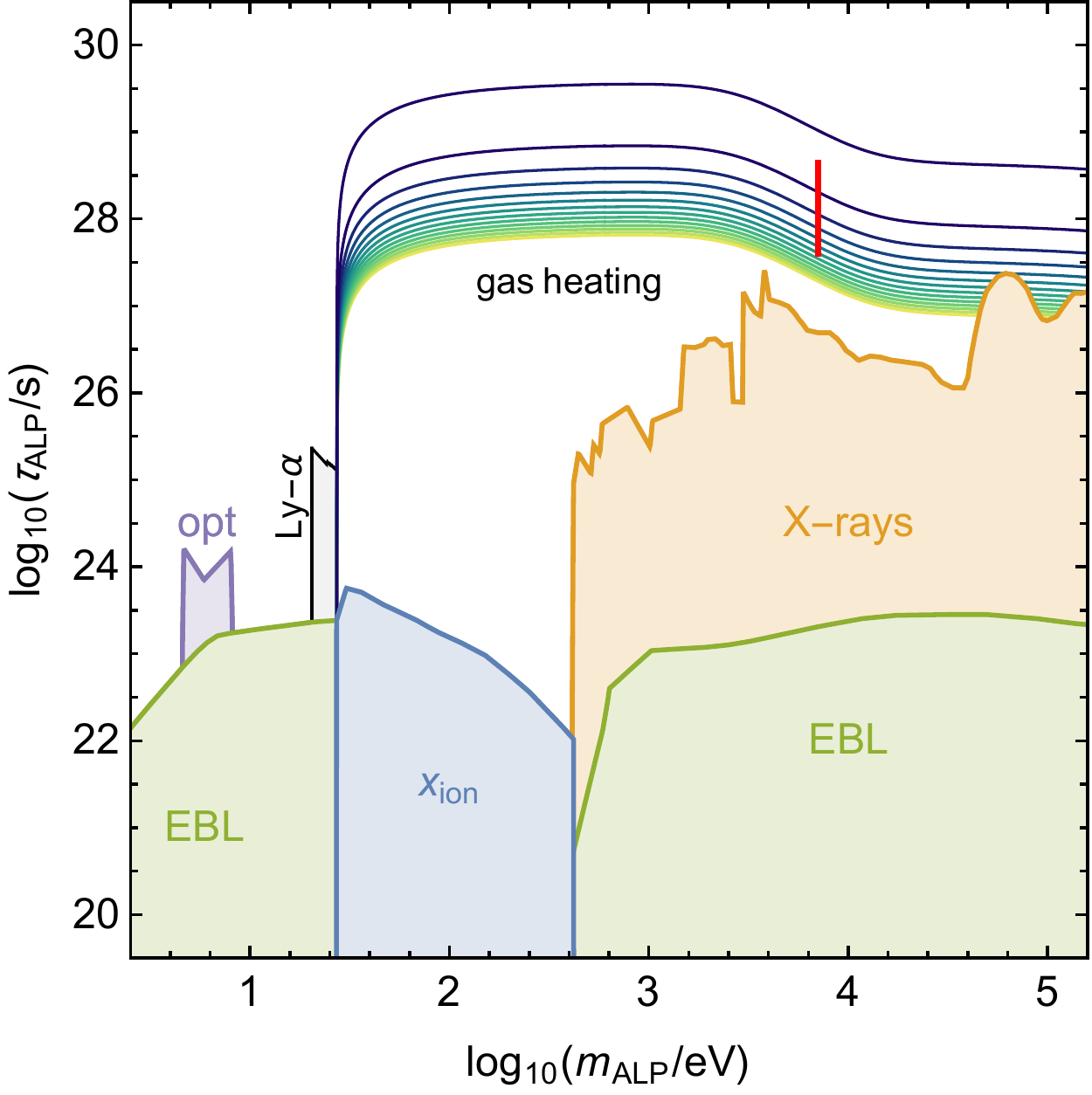} \hspace{4mm}
\includegraphics[height=0.45\textwidth]{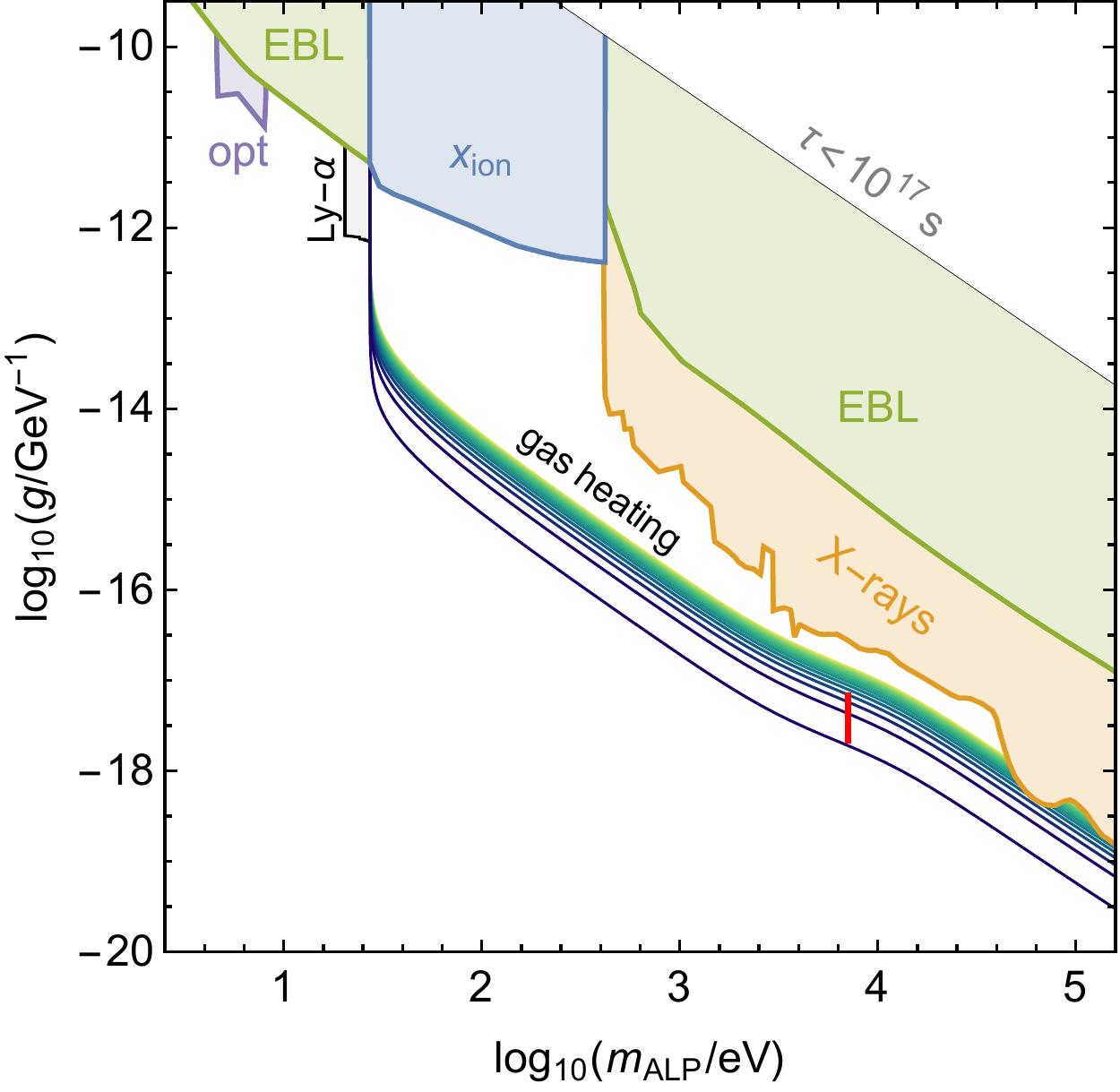}
\end{center}
\caption{The gray region represents the exclusion bound for the Lyman-$\alpha$ constraint of Eq.~\eqref{lacj1}. The colored lines represent instead the bound due to the gas heating assuming $T_k^{\rm max} = 7,8,9,\dots,20\,{\rm K}$ from blue to yellow, respectively. The other presented exclusion regions are taken from Ref.~\cite{Cadamuro:2011fd} and all the constraints are for evaluated $f_{\rm dm}=1$. The red line highlight the parameters for which the model reproduces the $3.5\,{\rm keV}$ signal assuming $f_{\rm dm}=1$.}
\label{fig:g_ALP}
\end{figure*}

\begin{figure*}[h!]
\begin{center}
\includegraphics[height=0.45\textwidth]{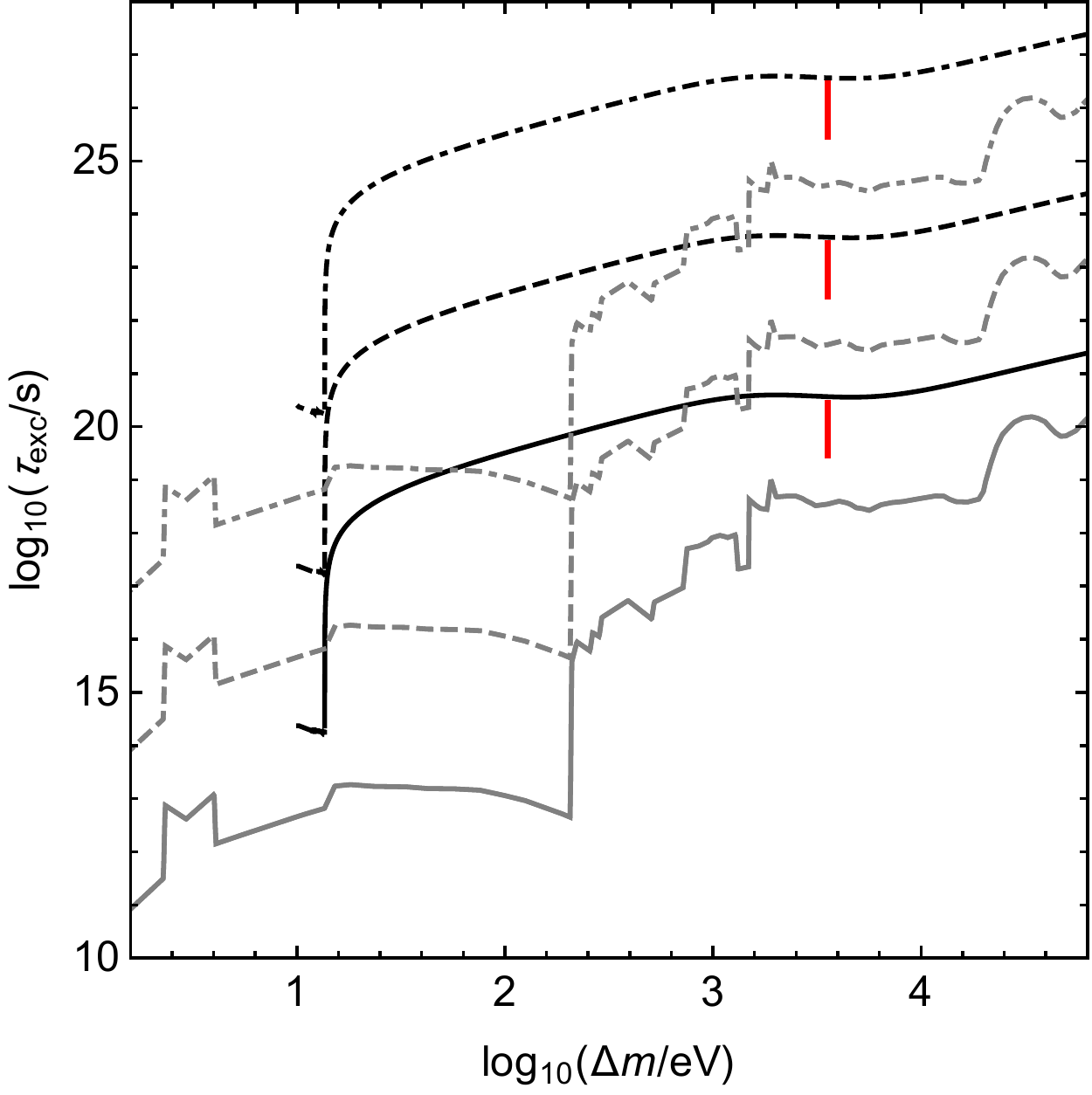} \hspace{4mm}
\includegraphics[height=0.45\textwidth]{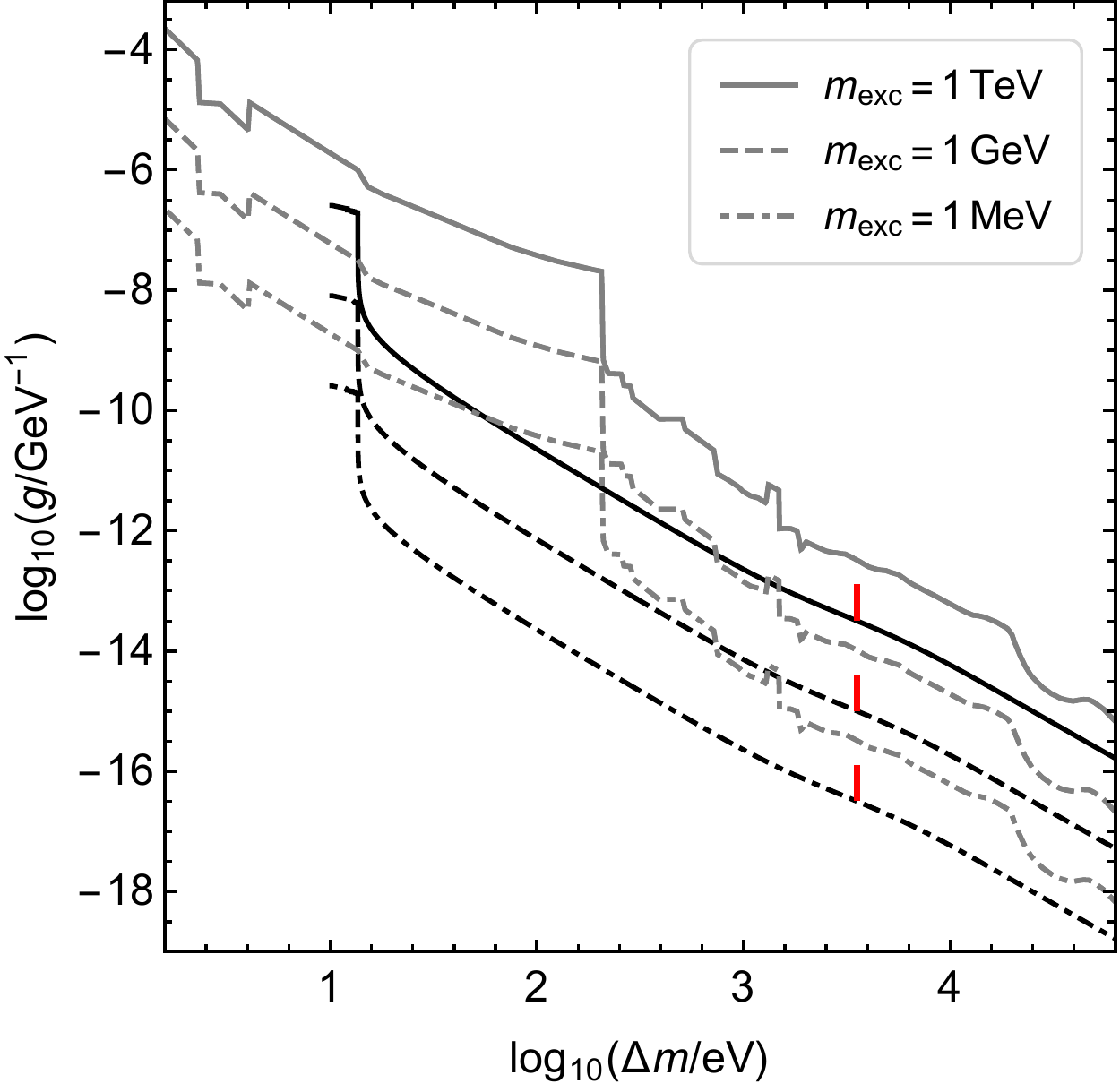}
\end{center}
\caption{The black lines show the bound from Lyman-$\alpha$ and gas heating (assuming $T_k^{\rm max}=7\,{\rm K}$) for different values of the DM particle mass $m_{\rm exc}$. The gray lines represents instead the constraints from Ref.~\cite{Cadamuro:2011fd} adapted to the case of exited DM. All the constraints are given for $f_{\rm dm}=1$. The red line highlight the parameters yielding a $3.5\,{\rm keV}$ emission compatible with observations for $f_{\rm dm}=1$.}
\label{fig:g_exc}
\end{figure*}

\section*{Results} 
\label{sec:Results}

Given the bound on the Ly-$\alpha$ intensity in Eq.~\eqref{lacj1} and on the photon emissivity, Eq.~\eqref{ghcj1}, we can constrain the properties of ALPs and excited DM that generate here the involved photon flux. The relevant processes are respectively the decay $X\to 2\gamma$, and the de-excitation $X^*\to \gamma X$, where $X$ is the DM particle. By defining $g$ as the effective coupling (with dimension of an inverse mass) responsible for these processes, we can write both the corresponding lifetimes as
\begin{equation}
\tau_{\rm dm} = \frac{8\pi}{g^2 E_*^3}\,,
\end{equation}
where $E_*$ is the energy of the emitted photon. In the ALP case $E_*=m_{\rm ALP}/2$, whereas for the excited DM case $E_*=\Delta m \equiv m_{X^*} - m_X$.

The results obtained for ALPs are shown in Fig.~\ref{fig:g_ALP}, under the assumption that they constitute the whole of the observed DM abundance. In the left panel we compare the lower bound on the ALP lifetime as a function of mass to the existing bounds taken from Ref.~\cite{Cadamuro:2011fd}. In the right panel we show instead the corresponding upper bound on the effective coupling $g$. The different lines represent different choices for the gas kinetic temperature. We remark that assuming the standard temperature for radiation at $z\sim17$, the EDGES measurement excludes at least at 95\% confidence level kinetic temperatures of the gaseous medium equal or larger than 7\,K~\cite{Hektor:2018qqw}.    

The case of excited DM is analysed in Fig.~\ref{fig:g_exc}, where we present the bounds on the lifetime of the excited state (left panel) and on the effective coupling (right panel), as a function of the mass splitting $\Delta m$. For the computation we assume that all DM is initially in the excited state. A rescaling factor of $\tau_{\rm exc} \sim f_{\rm dm}$ or $g\sim f_{\rm dm}^{-1/2}$ should then be applied if different scenarios are considered. The black lines refer to different choices of the DM mass, which we assume to be much larger than the mass splitting. The gray lines indicate instead the constraints inferred from ALP bounds.

The red lines in Fig.~\ref{fig:g_ALP} and Fig.~\ref{fig:g_exc} highlight the parameters for which the models reproduce the 3.5\,keV emission line~\cite{Bulbul:2014sua, Boyarsky:2014jta}. As these values fall within the exclusion region marked by the 7\,K contour, the EDGES measurement excludes the ALPs and excited DM explanations of the 3.5\,keV line at least at 95\% confidence level.

\section*{Discussion} 
\label{sec:Discussion}

Motivated by the recent EDGES measurement of the 21\,cm spectrum, in this Letter we have analysed the power of this observable to test a class of dark matter models resulting in photon injections into the gaseous medium after recombination. 

After deriving the limits implied by the EDGES observation on the intensity of Lyman-$\alpha$ photons and on the electron emissivity, we find that the corresponding bounds outperform the constraints previously considered in literature for Axion-like particle and excited dark matter on the masses range spanning from 20.4~eV to 0.1~MeV. We also find that assuming a soft photon background compatible with the $\Lambda$CDM prediction, the EDGES measurement rules out at 95\% confidence level the ALPs and excited dark matter explanation of the 3.5~keV signal detected mainly in galaxy clusters observations. 

Regardless of the origin of the anomalous absorption profile detected by the EDGES experiment, our work provides an important example of the reach of the 21\,cm spectrum observations.  

\section*{Acknowledgements} 
\label{sec:Acknowledgements}
We would like to thank Kristjan Kannike and Indrek Vurm for useful discussions.
This work was supported by the Estonian Research Council grants MOBTT5, IUT23-6, IUT26-2, PUT1026, PUT799, PUT808, the ERDF Centre of Excellence project TK133 and the European Research Council grant NEO-NAT.

\begin{appendices}

\section*{Appendix: corrections to the `on-spot' approximation}
The `on-spot' correction factor shown in Fig.~\ref{fig:MFP_f_param} is defined as the ratio of deposited to locally generated energy. Explicitly we have

\begin{equation}\label{app_eq1}
f(E,z)=\frac{4\pi\int \alpha(E,z)I_E(E,z)\,{\rm d}E}{j(z)}\,,
\end{equation}

where $j$ is the local photon emissivity. For the photon energies of interest,  $\sim$10-$10^5$~eV, the effective absorption coefficient $\alpha$ includes contributions from photoionization and from Compton scattering. The spectral intensity $I_E$ in Eq.~\eqref{app_eq1} is 

\begin{equation}
\begin{aligned}
I_E(E,z) &= \frac{1}{4\pi}\int_z^\infty\frac{{\rm d}z'}{H(z')(1+z')}\left(\frac{1+z}{1+z'}\right)^3
\\
& \quad\times j_E(z',E')e^{-\tau(E,z,z')}\,,
\end{aligned}
\end{equation}

where $E'=\frac{1+z'}{1+z}E$ and the optical depth

\begin{equation}
\tau(E,z,z')=\int_z^{z'} \frac{{\rm d}z''}{H(z'')(1+z'')}\alpha(E'',z'')\,.
\end{equation}

Since our spectral emissivity assumes the form

\begin{equation}
j_E(E,z)=j_0\delta(E-E_*)(1+z)^3\,,
\end{equation}

we can write $j(z)$ in Eq.~\eqref{app_eq1} as $j_0(1+z)^3$, thereby obtaining  for the spectral intensity at large $z$\footnote{Where the  contribution from $\Omega_\Lambda$ can be neglected.}
\begin{equation}
\begin{aligned}
	&I_E(E,z)
	=\frac{d_H}{4\pi\sqrt{\Omega_m}}j_0(1+z)^{3/2}\frac{E^{3/2}}{E_*^{5/2}}
	\\ & \times\exp\left[-\frac{d_H}{\sqrt{\Omega_m}}\hspace{-0.5cm}\int\limits_z^{(1+z)E_*/E-1}\hspace{-0.5cm}\frac{{\rm d}z'}{(1+z')^{5/2}}\alpha(E',z')\right]\,,
\end{aligned}	
\end{equation}
with $d_H$ being the Hubble distance.

\end{appendices}


\bibliographystyle{JHEP}
\bibliography{AB}

\end{document}